\documentclass[12pt,a4paper]{article}
\usepackage[utf8]{inputenc}
\usepackage[T1]{fontenc}
\usepackage{amsmath}
\usepackage{amsfonts}
\usepackage{amssymb}
\usepackage{graphicx}
\usepackage{hyperref}
\usepackage[numbers,sort&compress]{natbib}
\usepackage{authblk}

\title{Benchmarking Hadronic Models in Geant4 for Detector Simulations}
\author[1,2]{S.D.~Savenkov}
\author[1,2]{A.O.~Svetlichnyi}
\author[1,2]{I.A.~Pshenichnov}
\affil[1]{Institute for Nuclear Research of the Russian Academy of Sciences, Moscow, Russia}
\affil[2]{Moscow Institute of Physics and Technology, Dolgoprudny, Russia}

\newcommand{\firstauthoremail}{savenkov.sd@phystech.edu}
\newcommand{\lastauthoremail}{pshenich@inr.ru}

\begin{document}

\date{}

\maketitle

% Email block after title
\begin{center}
    \small
    Correspondence: \href{mailto:\firstauthoremail}{\firstauthoremail} (Alexandr Svetlichnyi), 
    \href{mailto:\lastauthoremail}{\lastauthoremail} (Igor Pshenichnov)
\end{center}

\begin{abstract}
The construction of modern detectors used in high-energy physics experiments is typically guided by modeling with the Geant4 toolkit to evaluate detector performance in terms of geometrical acceptance and detection efficiency. Several hadronic models are available in Geant4 for modeling nuclear reactions induced by fast nucleons. It is shown that they  result in different multiplicity and charge distributions of secondary particles and nuclear fragments.  The impact of these differences on modeling the Highly-Granular Neutron Detector (HGND) prototype for the BM@N experiment at Nuclotron-based Ion Collider fAcility (NICA) is evaluated.
\end{abstract}

\vspace{0.5cm}

\section{Introduction}\label{intro}

 The construction of sophisticated detectors employed in high-energy physics experiments is typically assisted by Geant4 toolkit~\cite{GEANT4:2002zbu,Allison:2006,Allison:2016lfl} modeling to evaluate the detector performance. Several sets of physics models of electromagnetic and hadronic processes are available in Geant4, and they are combined into Physics Lists~\cite{G4RPL} to be used in specific applications: nuclear and particle physics experiments, astrophysics and space science, medical physics and radiation protection~\cite{Allison:2016lfl}. Taking high-energy physics alone, at least three different Reference Physics Lists (RPL) are suggested for simulations~\cite{G4RPL}: FTFP\_BERT, QGSP\_BIC and QGSP\_INCLXX. As follows just from their naming scheme, the main difference between them consists in using different models of hadronic interactions of particles, light and heavy nuclei with nuclei in detector materials. The FTFP\_BERT RPL is usually considered as the default option for modeling in high-energy physics~\cite{Allison:2016lfl}. 

While it is reasonable to start a Geant4 modeling of a new detector with FTFP\_BERT, 
it is not a priory clear whether this default RPL will provide the most accurate results at all energies for all detector configurations. 
In this work, the light signals induced by high energy neutrons in the Highly-Granular Neutron Detector (HGND) prototype~\cite{Zubankov:2025abb} were calculated with three RPL which involve different hadronic intranuclear cascade models in simulations.  Presently, the HGND prototype is used in the BM@N experiment at NICA accelerator complex for detecting forward neutrons. Using Geant4 v11.2, we compare the predictions of the Bertini Cascade~\cite{Wright:2015xia}, Binary Light Ion Cascade~\cite{Folger:2004zma} and Liège IntraNuclear Cascade (INCL++)~\cite{Boudard:2002yn} models, implemented in the considered RPLs. The interaction of protons with the HGND prototype was also studied for the comparison of the detector response to neutrons.  The mass distributions of secondary nuclear fragments and their influence on the detector response were analyzed. 

\section{\label{sec:modeling} Hadronic models and RPL of Geant4}

Particle transport stands in the very core of any Monte-Carlo modeling of a detector. An initial primary particle interacts with the detector materials, deposits a part of its energy at each transportation step and produces secondary particles, including hadrons, which also contribute to the detector signal. Therefore, a reliable modeling of particles production in hadronic interactions is important for a realistic calculation of the detector performance, and a set of hadronic models is required for robust results. 

Three different RPLs were employed in this work for the modeling of the HGND prototype: FTFP\_BERT\_HP, QGSP\_BIC\_HP, and QGSP\_INCLXX\_HP.  These RPL also include the so-called high precision (HP) modeling of  nuclear reactions induced by low energy neutrons below 20~MeV.  In FTFP\_BERT\_HP, the Bertini intranuclear cascade model is used for reactions induced by neutrons and protons as well as mesons with energies below 6~GeV. A smooth transition to the FRITIOF model~\cite{Nilsson-Almqvist:1986ast} used above 6 GeV is implemented for these reactions within the 3-6 GeV interval. Inelastic nucleus-nucleus collisions are modeled using the Binary Light Ion Cascade (BIC) model in  FTFP\_BERT\_HP. In QGSP\_BIC\_HP, the BIC model is employed for simulating both nucleus-nucleus collisions and reactions induced by neutrons and protons, while meson interactions are still handled by the Bertini cascade model. In QGSP\_INCLXX\_HP, the INCL++ model is applied for proton- and neutron-induced reactions and nucleus-nucleus collisions except for the collisions where the mass number of both nuclei exceeds 18. While there are remaining differences between the considered RPLs, they do not affect the simulation results obtained in this work.

\section{\label{sec:model_spec} Detector configuration}

Following the general design of the HGND prototype~\cite{Zubankov:2025abb}, the detector configuration in simulations was represented by 14 modules, each containing an absorber layer and a scintillator layer. The transverse size of each module and, therefore, the detector, was taken as $12\times12$~cm$^2$. The thickness of each scintillator layer was set to 2.5~cm. Each of first five modules contained 8~mm thick lead absorbers, while each of the remaining nine modules contained 3~cm thick copper absorbers. The first module is preceded by an additional VETO scintillator layer. Since the transverse distributions of energy deposited in the HGND prototype were not considered in this work, each scintillator layer was modeled as a monolithic scintillator volume, in contrast to its transverse segmentation in the real detector. Printed circuit boards and support structures were also excluded to simplify the modeling focused on the longitudinal distributions of deposited energy along the main axis of the detector. 

Protons and neutrons with kinetic energies of 600~MeV and 3.8~GeV were transported along the main axis of the detector in simulations. The value of 3.8~GeV represents the average energy of spectator nucleons from fragmentation of $^{124}$Xe beam nuclei in the BM@N experiment~\cite{Zubankov:2025abb} and also the energy of neutrons from electromagnetic dissociation (EMD) of $^{124}$Xe~\cite{Pshenichnov:2024azq}. Kinetic energy of 600 MeV is typical for neutrons and protons emitted from a hot overlap zone in nucleus-nucleus collisions.

\section{\label{sec:results} Modeling results}

\subsection{\label{sec:res_mass} Charge distributions of secondary fragments}

\begin{figure}[htb!]
    \centering
    \includegraphics[width=0.9\linewidth]{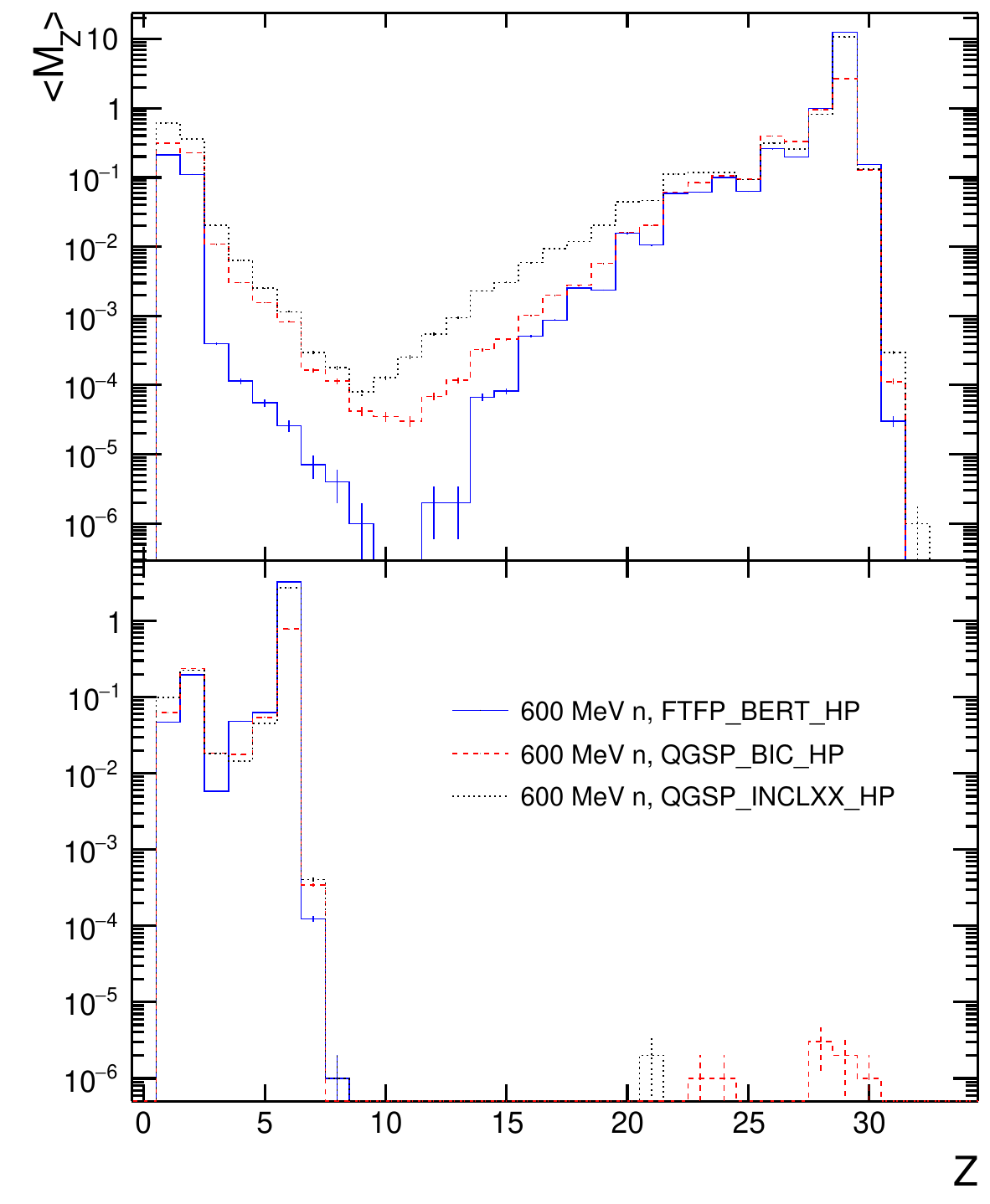}
    \caption{Average multiplicities of secondary fragments produced in all copper absorber  layers (top) and of fragments propagated to the adjacent scintillator layers (bottom) calculated with three RPLs for 600~MeV neutrons.}
    \label{fig:charge} 
\end{figure}

The average multiplicities of secondary charged fragments $\langle M_Z \rangle$ produced by 600~MeV neutrons in all copper absorber layers are presented in the top panel of Fig.~\ref{fig:charge}. The difference in $\langle M_Z \rangle$ obtained with different hadronic cascade models is evident for fragment charges $Z$ from 3 to 15. However, the difference between three calculation options is reduced for those secondary fragments which reach the scintillator layers, as seen in the bottom panel of Fig.~\ref{fig:charge}. This is explained by the absorption of heavy fragments with $Z>7$ inside the copper close to their production points, which makes improbable their propagation to the adjacent scintillator layers. As a result, only $Z\leq 7$ fragments produced in the absorber layers are responsible for the energy deposition in the scintillator layers along with $Z\leq 7$ fragments directly produced on C, N and O nuclei in the scintillator material.

The calculations with 3.8~GeV neutrons with three RPLs demonstrate a less prominent difference with respect to produced fragments. It is also reduced in the scintillator layers. The same conclusions are valid for lead absorbers and primary protons. 

\subsection{\label{sec:res_signal} Signals in scintillator layers}

Following the calculation of energy deposited by charged fragments in each of 14 scintillator layers, the nonlinearity of the light yield function according to the Birks' law was taken into account. In addition, 
the signal smearing in detecting light by the photomultiplier tubes (PMT) was also considered assuming that the light signals were  collected within a typical time window of 300~ns.

\begin{figure}[htb!]
    \centering
    \includegraphics[width=0.9\linewidth]{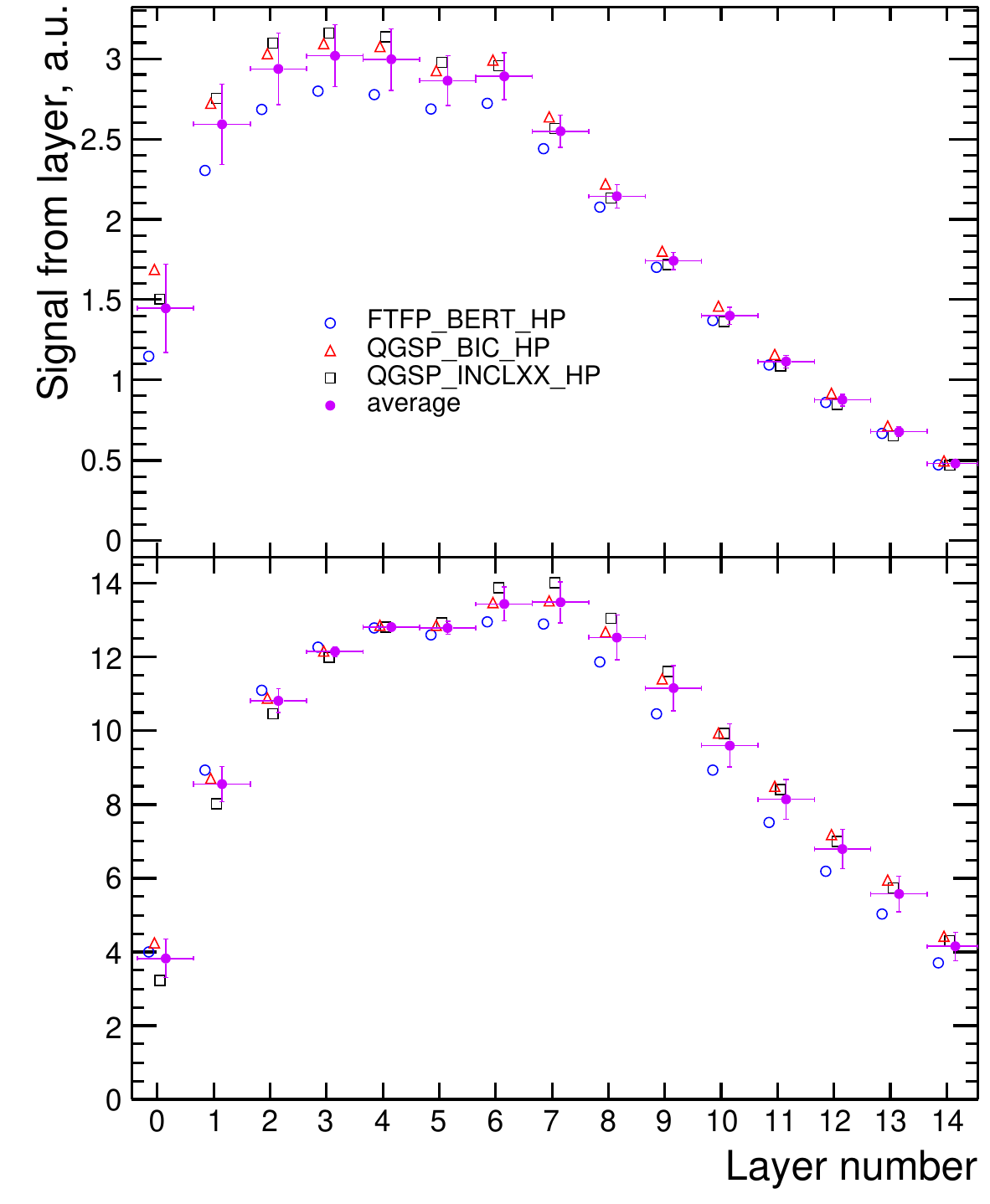}
    \caption{Signals in scintillator layers from 600~MeV (top) and 3.8~GeV (bottom) neutrons calculated with three RPLs. The layer number 0 is the VETO layer. Points with error bars represent the mean values of signals calculated with three options and their standard deviations.}
    \label{fig:signal}
\end{figure}

In Fig.~\ref{fig:signal} signals in each layer are presented. While the modeling with QGSP\_BIC\_HP and QGSP\_INCLXX\_HP shows similar results, the involvement of FTFP\_BERT\_HP results in lower response in certain layers. For 600~MeV neutrons the largest difference between the models is observed in the scintillator layers adjacent to lead absorbers. In contrast, the results for 3.8~GeV neutrons noticeably differ specifically in the scintillator layers adjacent to copper absorbers. In addition to the signals calculated with specific RPLs, the mean values of these signals are also presented in Fig.~\ref{fig:signal} together with respective standard deviations to be considered as systematic simulation uncertainties. These systematic uncertainties are equal to 19\%(14\%) in the first layer, 6\%(4\%) in the layer with the highest signal and 3\%(9\%) in the last layer for neutrons with energy of 600~MeV(3.8~GeV).

\section{\label{sec:discussion} Conclusion}

The interactions of 600~MeV and 3.8~GeV neutrons and protons in the HGND prototype made of alternate absorber and scintillator layers were simulated with FTFP\_BERT\_HP, QGSP\_BIC\_HP and QGSP\_INCLXX\_HP Reference Physics Lists of Geant4 v11.2 toolkit. As found, all three simulation options provide different average multiplicities of secondary fragments produced in absorber materials (copper or lead). However, only $Z\leq 7$ fragments propagate further to adjacent scintillator layers, and they are mixed with $Z\leq 7$ fragments directly produced on C, N and O nuclei of scintillator material. As a result, a better agreement between the multiplicity distributions of $Z\leq 7$ fragments in scintillator layers calculated with different RPLs was found. This results to a reduced difference  between the light signals from scintillator layers calculated with different RPLs. The systematic  uncertainties associated with the choice of hadronic cascade models of Geant4 toolkit are equal to 19\%(14\%) in the first layer, 6\%(4\%) in the layer with the highest signal and 3\%(9\%) in the last layer for neutrons with energy of 600~MeV(3.8~GeV). These systematic uncertainties contribute to the uncertainties in calculating the efficiency of the HGND prototype. Further modeling with detailed geometry of the detector and its trigger is planned as well as the comparison of calculated spatial distributions of detector response with BM@N data.

\section*{Acknowledgements}
The authors are grateful to the members of the HGND team of the BM@M Collaboration for fruitful discussions concerning the detector geometry. The work is supported by the Ministry of Science and Higher Education of the Russian Federation, Project FFWS-2024-0003.

\bibliographystyle{elsarticle-num}

\bibliography{benchmarking-models-g4}

\end{document}